\documentclass[aoas,nameyear,dvips]{arximspdf}
\usepackage{graphicx}

\doi{10.1214/10-AOAS369}
\volume{5}
\issue{1}
\pubyear{2011}
\firstpage{150}
\lastpage{175}

\makeatletter
\newcommand{\E}{\mathrm{E}}
\newcommand{\var}{\operatorname{Var}}
\newcommand{\cN}{\mathcal{N}}
\newcommand{\bmu}{\bolds{\mu}}
\newcommand{\I}{\mathbf{I}}

\newcommand{\diag}{\operatorname{diag}}

\newcommand{\balpha}{\bolds{\alpha}}
\newcommand{\bbeta}{\bolds{\beta}}
\newcommand{\btau}{\bolds{\tau}}
\newcommand{\bSigma}{\bolds{\Sigma}}

\newcommand{\A}{{\mathbf{A}}}
\newcommand{\B}{{\mathbf{B}}}
\newcommand{\bfT}{{\mathbf{T}}} 
\newcommand{\V}{{\mathbf{V}}}
\newcommand{\X}{{\mathbf{X}}}
\newcommand{\Y}{{\mathbf{Y}}}

\newcommand{\0}{{\mathbf{0}}}
\newcommand{\h}{\mathbf{h}}

\makeatother

\begin{document}
\begin{frontmatter}

\title{A spatial analysis of multivariate output from regional climate models}
\runtitle{Spatial analysis of RCMs}

\begin{aug}
\author[A]{\fnms{Stephan R.} \snm{Sain}\corref{}\ead[label=e1]{ssain@ucar.edu}\thanksref{aut1}},
\author[B]{\fnms{Reinhard} \snm{Furrer}\ead[label=e2]{reinhard.furrer@math.uzh.ch}\thanksref{aut2}}
and
\author[C]{\fnms{Noel} \snm{Cressie}\ead[label=e3]{ncressie@stat.osu.edu}\thanksref{aut3}}
\runauthor{S. R. Sain, R. Furrer and N. Cressie}
\thankstext{aut1}{Supported in part by NSF Grants DMS-07-07069 and ATM-05-34173.}
\thankstext{aut2}{Supported in part by NSF Grant DMS-06-21118. In addition, much
of the research in this paper was done while the second author was an assistant
professor in the Department of Mathematical and Computer Sciences at the Colorado School
of Mines in Golden, CO.}
\thankstext{aut3}{Supported in part by the Office of Naval Research under Award N00014-08-1-0464.}
\affiliation{National Center for Atmospheric Research, University of
Zurich and \mbox{Ohio State University}}

\address[A]{S. R. Sain\\
Geophysical Statistics Project\\
Institute for Mathematics Applied to Geosciences\\
National Center for Atmospheric Research\\
P.O. Box 3000, Boulder\\
Colorado 80307\\
USA\\
\printead{e1}} 

\address[B]{R. Furrer\\
Institute of Mathematics\\
University of Zurich\\
Zurich\\
Switzerland\\
\printead{e2}}

\address[C]{N. Cressie\\
Department of Statistics\\
Ohio State University\\
Columbus, Ohio 43210\\
USA\\
\printead{e3}}

\end{aug}

\received{\smonth{10} \syear{2009}}
\revised{\smonth{6} \syear{2010}}

%
\begin{abstract}
Climate models have become an important tool in the study of climate
and climate change, and ensemble experiments consisting of multiple
climate-model runs are used in studying and quantifying the uncertainty
in climate-model output. However, there are often only a limited number
of model runs available for a particular experiment, and one of the
statistical challenges is to characterize the distribution of the model
output. To that end, we have developed a multivariate hierarchical
approach, at the heart of which is a new representation of a
multivariate Markov random field. This approach allows for flexible
modeling of the multivariate spatial dependencies, including the
cross-dependencies between variables. We demonstrate this statistical
model on an ensemble arising from a regional-climate-model experiment
over the western United States, and we focus on the projected change in
seasonal temperature and precipitation over the next 50 years.
\end{abstract}

%

\end{frontmatter}

\section{Introduction}

Many processes in the Earth system cannot be directly observed, and
computer modeling has become a primary mode for studying these
processes. These computer models often encapsulate entire fields of
knowledge, providing a virtual laboratory for understanding physical
relationships and serving as a basis for making predictions. The
Earth's climate, for example, is determined by the flows of energy,
water, gases, etc., within and between the different components of the
climate system, including atmosphere, oceans, terrestrial and marine
biospheres, sea ice, etc. Climate models attempt to represent this
system, as well as to incorporate anthropogenic forcings to assess the
impact of human intervention.

While climate models are deterministic, the output generated by these
models is complex and subject to a number of sources of uncertainty.
Initial climate states, assumptions about future forcings, and, of
course, our understanding (or lack thereof) of the physical processes
and their representation in computer models, are all issues that may
lead to uncertainty in the model output.
To gain a better understanding of this model uncertainty, there is an
increasing use of ensembles consisting of multiple model runs. These
experiments may involve varying initial conditions (simple ensembles),
model physics (perturbed-physics ensembles), specific models
(multi-model ensembles), or some combination of all three, in an
attempt to capture the range of variation in the model output.

Climate is often considered the long-term average (or, more generally,
the long-term distribution) of weather, and climate models typically
simulate decades of weather to capture this long-term behavior. These
simulations can take weeks or even months of computing time on
high-performance supercomputers. Furthermore, as interest continues to
grow in climate models with higher spatial resolution, in particular,
to study regional and local impacts of climate and climate change, the
computational demands of climate modeling continue to grow. Even with
the increased use of ensembles, the number of ensemble members is
typically limited due to constraints on the models available,
computation, funding, etc. Hence, statistical methods are necessary to
quantify the distribution and breadth of variation of the model output
in the ensemble. To this end, we introduce a hierarchical statistical
model to capture the multivariate spatial distribution of the output
fields (e.g., the joint spatial distribution of temperature and
precipitation) from a regional-climate-model ensemble. With this
statistical representation of the model output, we can present
probabilistic projections of regional climate change based on the
ensemble. Furthermore, by considering multiple output fields
simultaneously, we can incorporate correlations between these fields,
improving joint projections necessary for many climate-impacts studies
(e.g., agriculture, water management, public health, etc.).

\subsection{Regional climate models}

The climate of a region is determined by processes that exist at
planetary, regional, and local spatial scales and across a wide range
of temporal scales (multi-decadal to sub-daily). This creates serious
difficulties when attempting to construct computer models that can
simulate regional climate. Atmosphere--ocean general circulation models
(AOGCMs or, more simply, GCMs) couple an atmospheric model with an
ocean model and seek to simulate the Earth's global climate system. Due
to model complexity, the need to simulate climate over decadal and even
centennial time scales, and computational limitations, these models
typically have grid boxes on the spatial scale of 200 to 500~km. While
these models are extremely useful for investigating the large-scale
circulation and forcings that affect the Earth's global climate, there
are limitations to their use for regional and local projections that
might be of interest to the climate-impacts community.

Recognizing the need to include large-scale processes, even when
studying regional climate, as well as the ability of GCMs to capture
such phenomena, there is considerable attention on developing
downscaling methodologies.
Downscaling refers simply to generating information on the basis of a
GCM, but at spatial scales below that of the GCM. There are two main
types of downscaling, dynamical and statistical. Statistical
downscaling is a computationally efficient approach that uses empirical
relationships to connect the coarse-resolution GCM output to regional
and local variables. This approach needs fine-scale and
long-time-duration observational data, and there is some uncertainty
about the stability of these empirical relationships over long periods
of time, especially with varying forcings (e.g., increasing CO$_2$
concentrations).

Dynamical downscaling involves using high-resolution climate models. Of
course, there are limitations due to computational demands and a price
to be paid for the higher resolution. One approach uses only the
atmospheric component of a GCM with, for example, observed ocean
temperatures. These so-called time-slice experiments are globally
consistent, but they generally use many of the same formulations as the
coarse-scale GCMs.

Regional climate models (RCMs) are the focus of this research and are
another dynamical approach based on high-resolution climate models.
These models typically focus on a limited spatial domain, have grid
boxes on the scale of 20 to 100~km, and there are often simplifications
of ocean processes in these models. They also use initial conditions
and time-dependent lateral boundary conditions from the GCM (e.g.,
winds, temperature, and moisture). Hence, global circulation and
large-scale forcings are consistent with the GCM, but, with the
higher-resolution forcings included (e.g., topography, land cover,
etc.), these models have the potential to actually enhance the
simulation of climate on regional and local scales. Of course, RCMs can
be influenced by potential biases in the GCM, and there is a lack of
two-way interactions between the driving GCM and the RCM.

For those interested in understanding more about climate and climate
change and global and regional climate modeling, we refer them to the
Intergovernmental Panel on Climate Change (IPCC, \url{http://www.ipcc.ch})
assessment reports, and, in particular, to the contributions of Working
Group I to the Third Assessment Report [\citet{IPCC2001}] and the
Fourth Assessment Report [\citet{IPCC2007}]. These documents not only
include excellent overviews of the issues but also numerous scientific
references for more in-depth coverage.

\subsection{A statistical representation of climate-model output}

We propose a statistical model for combining the output from simple
ensembles of RCMs in order to characterize the distribution of the
model output. This statistical model will be formulated through what
has now become the standard three-level hierarchical formulation,
namely, data model, process model, and parameter model (prior
distribution). The data model links the RCM output to an unobserved
spatial process, where this process model is formulated to capture the
spatial variation in the RCM output. Both the data model and the
process model depend on unknown parameters to which a prior
distribution is assigned.

This basic hierarchical approach has been used in other settings for
combining climate-model output. For example, \citet{Teb05} focuses on
univariate summaries of temperature change from an ensemble of GCMs.
\citet{SmithTebaldi2008} also explore univariate summaries of
temperature change from an ensemble of GCMs and link these summaries
over different regions around the globe. \citet{Tebaldi08} extend
these approaches to bivariate models of temperatures and precipitation.
\citeauthor{Fur07} (\citeyear{Fur07,furrersain2007}) study univariate spatial summaries and
\citet{BerlKim08} study bivariate time series, again constructed from
an ensemble of GCMs. More recently, \citet{Cooley2010} and \citet
{Kang2010} have applied hierarchical models to ensembles of RCMs.
However, to our knowledge, this paper is the first approach of this
kind for a \textit{spatial} analysis of \textit{multivariate} output
from RCMs.

At the heart of this statistical model is an implementation of a
multivariate Markov random field (MRF) for lattice data that offers
great flexibility in modeling the spatial cross-dependencies between
variables. We emphasize this capability in this setting, as the
underlying physical behavior of the climate system suggests the
potential for significant spatial cross-dependencies, in particular,
for key variables like \textit{temperature} and \textit
{precipitation}. While
these two output fields are the focus of this work, we note that the
basic approach based on the multivariate MRF presented here can be
easily modified to consider other output fields as well.

MRF models are also excellent tools for analyzing data laid out on
regular spatial lattices,
such as those associated with images, remote-sensing, climate models,
etc., or on irregular spatial lattices, such as U.S. census divisions
(counties, tracts, or block-groups) or other administrative units. In
contrast to geostatistical methods that model spatial dependence
through the specification of a covariance function (typically based on
distances between spatial locations), Gaussian MRF models represent the
conditional expectation of an observation at a spatial location as a
linear combination of observations at neighboring locations. Spatial
dependence is induced through this conditional autoregression and the
choice of neighborhoods.

In addition, using a MRF formulation will allow us to incorporate
computational advantages due to the gridded nature of the climate-model
output and the sparseness that is characteristic of the
spatial-precision matrices (inverse covariance matrices) that are
specified in such models. Furthermore, the multivariate nature of the
statistical model (more than one model output considered at each grid
box) will allow for more complex inferences that are of use to those
studying impacts of climate and climate change.

\subsection{Outline}

In Section \ref{sec:mrf} an overview of MRF models is presented,
followed by a description of the new formulation in Section \ref
{sec:new}. Section \ref{sec:hier} contains the details of our
hierarchical specification. An extensive study of an application using
a simple ensemble of RCM output, focusing on changes in seasonal
temperature and precipitation, will be presented in Section \ref
{sec:ex}. 
Concluding remarks are given in Section \ref{sec:conclude}.

\section{MRF and CAR models}
\label{sec:mrf}

Besag (\citeyear{Besa74}) laid out the basic framework for MRF models. For random
variables $y_1,\ldots,y_n$ observed at $n$ locations on a
spatial-lattice structure, the collection of conditional distributions
$f(y_i|y_{-i}),\break i=1,\ldots, n$ (where $y_{-i}$ refers to all random
variables except the $i$th one), can be combined under certain
regularity conditions to form a joint distribution $f(y_1,\ldots,
y_n)$. \citet{RueHeld05} can be consulted for an excellent exposition
of the theory of MRFs; see also the reviews in the texts by \citet
{Cres93}, \citet{BaneCarlGelf04}, and \citet{SchaGotw05}.
Conditional autoregressive (CAR) models are special cases of MRF models
where the conditional distributions are assumed to be Gaussian.

\subsection{Univariate CAR models}
\label{sec:mrf1}

In the univariate setting and assuming Gaussian conditional
distributions for $f(y_i|y_{-i})$, the conditional mean and conditional
variance associated with $f(y_i|y_{-i})$ are specified as
\[
\E[y_i|y_{-i}] = \mu_i + \sum_{j\neq i}^n b_{ij} (y_j-\mu_j)
\quad \mbox{and}\quad
\var[y_i|y_{-i}] = \tau_i^2>0,
\]
where $b_{ii}=0$; $i=1,\ldots, n$. 
Under regularity conditions, this collection of conditional
distributions gives rise to a joint Gaussian distribution,
%
\begin{equation}\label{unijoint}
\cN\bigl(\bmu,(\I-{\B})^{-1}\mathbf{T} \bigr),
\end{equation}
where $\bmu=(\mu_1,\ldots,\mu_n)^\prime$, $\I$ is an $n\times n$
identity matrix, $\B$ is the $n\times n$ matrix with the ($i,j$)th
element $b_{ij}$, and $\mathbf{T} = \diag(\tau_1^2,\ldots,\tau_n^2)$. The
regularity conditions are on the spatial-dependence parameters, $\{
b_{ij}\}$, and they ensure that the resulting matrix, $(\I-{\B
})^{-1}\mathbf{T}$, is a \textit{bona fide} covariance matrix; that is,
$(\I-\B)^{-1}\mathbf{T}$ is symmetric and positive-definite.
The spatial dependence is induced by the autoregression, which is
determined by setting the coefficients
$b_{ij}\neq0$ if $j \in N_i$ (and $0$ otherwise), where $N_i$ is a
collection of indices that define a neighborhood of the $i$th location
in the spatial lattice.


\subsection{Multivariate CAR models}
\label{sec:old}

Mardia (\citeyear{Mard88}) extended the MRF model of \citet{Besa74} to the
multivariate setting where there is more than one measurement at each
lattice point. In particular, let $\mathbf{y}_i$ be a
$p$-dimensional random
vector. Then, for $i=1,\ldots, n$, let $f(\mathbf{y}_i|\mathbf{y}_{-i})$ be a Gaussian
distribution of $\mathbf{y}_i$, given all random vectors
except the $i$th, with
\[
\E[\mathbf{y}_i|\mathbf{y}_{-i}] = \bmu_i +
\sum_{j\neq i} \B_{ij}(\mathbf{y}_j - \bmu_j)
\quad \mbox{and}\quad
\var[\mathbf{y}_i|\mathbf{y}_{-i}] = \mathbf{T}_i,
\]
where $\bmu_i$ is a $p$-dimensional vector, $\B_{ij}$ is a 
$p\times p$ matrix, and $\mathbf{T}_i$ is a $p\times p$ covariance matrix.
Assume that $\B_{ij}\mathbf{T}_j = \mathbf{T}_i\B_{ji}^\prime$, for all
$i,j=1,\ldots,n$ (to ensure symmetry). Further, assume $\B_{ii} = -\I$
and $\B_{ij}\neq\mathbf{0}$ for $j\in N_i$ and $i=1,\ldots,n$.
Under the assumption that 
the $np\times np$ matrix $\hbox{Block}(-\B_{ij})$ (i.e., a block matrix
with blocks given by $-\B_{ij}$) is positive-definite,
\citet{Mard88} establishes that $\mathbf{y}= (\mathbf{y}_1^\prime,\ldots,\mathbf{y}_n^\prime
)^\prime$ follows a
$\cN(\bmu,\bSigma)$ distribution where
$\bmu= (\bmu_1^\prime,\ldots,\bmu_n^\prime)^\prime$ and $\bSigma=
(\hbox{Block}(-\mathbf{T}_i^{-1}\B_{ij}) )^{-1}$.

As written, this formulation is overparameterized and there have been a
number of efforts in the literature that focus on ways of specifying
the parameters in the basic model. See, for example, \citet{Bill97},
\citet{Kim01}, 
\citet{Pett02}, \citet{CarlBane03}, \citet{GelfVoun03}, \citet
{Jin2005}, \citet{Jin2007}, \citet{Dan2006}, \citet{SainCres07}, among others.

\section{An alternative formulation of a multivariate MRF}
\label{sec:new}

It is our experience that the basic multivariate MRF model of \citet
{Mard88} is difficult to implement in practice without dramatic
simplification of the matrices representing the spatial dependence
parameters [e.g., \citet{Bill97}] or the use of restrictive priors on
the elements of these same matrices [\citet{SainCres07}].
Here, we fully develop a new way of representing multivariate lattice
data that was suggested, but not implemented, by \citet{SainCres07},
for the purposes of analyzing an RCM experiment.

The multivariate extension of the framework laid out by \citet{Besa74}
and explored by \citet{Mard88} is based on the assumption of a
multivariate observation at each point on a standard two-dimensional
spatial lattice. Fundamental to our approach is 
thinking of multivariate lattice data as univariate data on a more
complex lattice structure.

\begin{figure}[b]

\includegraphics{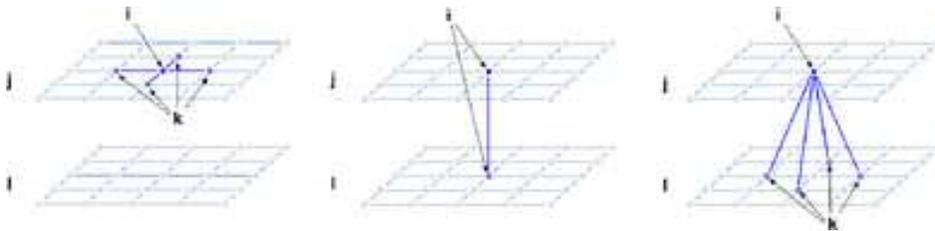}

\caption{Examples of different types of neighborhoods. The left frame
shows a within-variable spatial neighborhood, while the middle frame
shows a within-location neighborhood. The right frame demonstrates the
neighborhood associated with cross-variable connections. See also
equation (\protect\ref{ev}).}
\label{neigh}
\end{figure}

In particular, this more complex lattice structure is conceptualized as
a ``stacking'' of the lattices associated with each variable.
Neighborhoods are defined by connections between locations for each
variable within a lattice and again for locations across each lattice
structure. A two-dimensional example of these different types of
neighborhoods is shown in Figure \ref{neigh}. A within-variable
dependence structure is induced by connecting locations within a
lattice associated with a particular variable (left frame).
Cross-dependencies, both within a location (middle frame) and across
locations (right frame), are induced through connections between the
lattices for different variables.

The key feature of this approach is that it still falls within the
original univariate framework of \citet{Besa74} outlined in Section
\ref{sec:mrf1}. Let $y_{ij}$ denote the $j$th variable observed at the
$i$th location on the lattice. Then, for each Gaussian conditional
distribution, the mean and variance need to be specified. With sums on
the right-hand side corresponding to specific types of neighborhoods in
Figure \ref{neigh}, the conditional mean is given by
%
\begin{eqnarray}\label{ev}
\E\bigl[y_{ij} | y_{-\{ij\}}\bigr] &=& \mu_{ij} + \sum_{k\neq
i}b_{ijkj}(y_{kj}-\mu
_{kj})\nonumber\\[-8pt]\\[-8pt]
&&{} + \sum_{\ell\neq j} b_{iji\ell}(y_{i\ell}-\mu_{i\ell}) +
\sum_{k,\ell\neq i,j} b_{ijk\ell}(y_{k\ell}-\mu_{k\ell})\nonumber
\end{eqnarray}
with the conditional variance given by
$\var[y_{ij} | y_{-\{ij\}}] = \tau_{ij}^2$,
for all lattice points $i=1,\ldots, n$, variables $j=1,\ldots, p$, and
with $y_{-\{ij\}}$ denoting all components of $\mathbf{y}$
except for the ($i,j$)th.

In the conditional mean, the coefficients in the first summation, $\{
b_{ijkj}; i =1,\ldots, n, k\in N_i, j=1,\ldots, p\}$, represent
connections within a particular layer and control conditional
dependence between the $i$th lattice point and neighboring points for
the $j$th variable (left frame in Figure \ref{neigh}).
The coefficients in the second summation, $\{b_{iji\ell}; i=1,\ldots,
n, j\neq\ell= 1,\ldots, p\}$, represent connections across layers at
the same lattice point and control conditional dependence between
variables $j$ and $\ell$ at the $i$th lattice point (middle frame in
Figure \ref{neigh}).
Finally, the coefficients in the third summation, $\{b_{ijk\ell};
i=1,\ldots, n, k\in N_i, j\neq\ell= 1, \ldots, p\}$, represent
connections between locations across layers for different variables and
control conditional cross-spatial dependence (right frame in Figure
\ref
{neigh}). Of course, all of these conditional-dependence parameters and
the variances $\{\tau_{ij}^2\}$ must satisfy regularity conditions that
yield a symmetric, positive-definite covariance matrix for the joint
distribution.

Some simplification of this basic structure is necessary to reduce the
dimensionality of the parameter space. Ordering the data as $\mathbf{y}=[\mathbf{y}
_1^\prime, \ldots, \mathbf{y}_n^\prime]^\prime$, where
$\mathbf{y}_i = [y_{i1},\ldots
,y_{ip}]^\prime$ represents variables at the $i$th lattice point, we
see from (\ref{unijoint}) that the joint distribution is Gaussian with
mean given by $\bmu=[\bmu_1^\prime,\ldots,\bmu_n^\prime]^\prime$ and
$\bmu_i=[\mu_{i1},\ldots,\mu_{ip}]^\prime$, and with an $np\times np$
covariance matrix given by
%
\begin{equation} \label{cov1}
\left[
\matrix{
\A_1 & \B_{12}\delta_{12} & \cdots& & \B_{1n}\delta_{1n} \cr
\B_{21}\delta_{21} & \A_2 & & & \vdots\cr
\vdots& & \ddots& \cr
& & & \A_{n-1} & \B_{n-1,n}\delta_{n-1,n}\cr
\B_{n1}\delta_{n1} & \cdots& & \B_{n,n-1}\delta_{n,n-1} & \A_n
}\right]^{-1}
\bfT,
\end{equation}
where $\delta_{ik} = 1$ if $k \in N_i$ and $0$ otherwise. Each $p
\times p$ block is given by
\[
\A_i = \left[
\matrix{
1 & & -b_{iji\ell} \cr
& \ddots\cr
-b_{i\ell ij} & & 1
}
\right] \quad \mbox{or}\quad
\B_{ik} = \left[
\matrix{
-b_{i1k1} & & -b_{ijk\ell} \cr
& \ddots\cr
-b_{i\ell kj} & & -b_{ipkp}
}
\right],
\]
where $-b_{iji\ell}$ and $-b_{i\ell ij}$ are arbitrary off-diagonal
elements of $\A_i$, and $-b_{ijk\ell}$ and $-b_{i\ell kj}$ are
arbitrary off-diagonal elements of $\B_{ik}$.
Finally, $\bfT=\diag(\tau_{11}^2,\break\dots, \tau_{1p}^2, \ldots,
\tau_{n1}^2,\ldots, \tau_{np}^2)$.

In general, we shall assume that the neighborhood structure is
symmetric; that is, if the $k$th lattice point is a neighbor of the
$i$th lattice point, then the $i$th lattice point is a neighbor of the
$k$th ($\delta_{ik}=\delta_{ki}$). We shall also assume that $\tau
_{ij}^2 = \tau_j^2$ for all $j$, implying a separate variance for each
variable that does not vary with location. Hence, $\bfT= \I_n \otimes
\diag(\btau)$, where $\btau= [\tau_1^2, \ldots, \tau_p^2]^\prime$.
Further assumptions on the spatial-dependence parameters are made to
reduce the dimensionality of the parameter space.

To address symmetry and variance homogeneity across location, it
suffices to examine the components of specific blocks in the inverse of
the covariance matrix given by (\ref{cov1}). First, the \textit{diagonal
blocks} are given by
\[
\operatorname{diag}(\btau)^{-1} \A_i
=
\left[
\matrix{
1/\tau_1^2 & & -b_{iji\ell}/\tau_j^2 \cr
& \ddots\cr
-b_{i\ell ij}/\tau_\ell^2 & & 1/\tau_p^2
}
\right].
\]
By symmetry, the corresponding off-diagonal elements should be equal;
that is,
$b_{iji\ell}/\tau_j^2 = b_{i\ell ij}/\tau_\ell^2$.
Setting $b_{iji\ell} = \rho_{j\ell}\tau_j/\tau_\ell$, with $\rho
_{\ell
j}=\rho_{j\ell}$, is one way of achieving the desired result. Then,
$\operatorname{diag}(\btau)^{-1} \A_i = \diag(\btau)^{-1/2}\A\diag(\btau
)^{-1/2}$,
where
\[
\A= \left[
\matrix{
1 & & -\rho_{j\ell} \cr
& \ddots\cr
-\rho_{j\ell} & & 1
}
\right].
\]

For the \textit{off-diagonal blocks}, symmetry demands that
%
\begin{equation}\label{Bcond}
\operatorname{diag}(\btau)^{-1}\B_{ik} = [\operatorname{diag}(\btau)^{-1}\B
_{ki}]^\prime.
\end{equation}
Assuming $i>k$, the left-hand side of (\ref{Bcond}) is given by
\[
\operatorname{diag}(\btau)^{-1} \B_{ik}
=
\left[
\matrix{
-b_{i1k1}/\tau_1^2 & & -b_{ijk\ell}/\tau_j^2 \cr
& \ddots\cr
-b_{i\ell kj}/\tau_\ell^2 & & -b_{ipkp}/\tau_p^2
}
\right].
\]
Likewise, setting $b_{ijkl} = \phi_{jl}\tau_j/\tau_\ell$ gives
$\operatorname{diag}(\btau)^{-1} \B_{ik} = \diag(\btau)^{-1/2} \B\times\break\diag
(\btau)^{-1/2}$,
where
\[
\B= \left[
\matrix{
-\phi_{11} & & -\phi_{j\ell} \cr
& \ddots\cr
-\phi_{\ell j} & & -\phi_{pp}
}
\right],
\]
%
and, for $i>k$, set
$\operatorname{diag}(\btau)^{-1}\B_{ki} = \diag(\btau)^{-1/2} \B^\prime
\diag
(\btau)^{-1/2}$ to satisfy (\ref{Bcond}).


The covariance in (\ref{cov1}) then simplifies to
%
\begin{equation}\label{cov2}
[\I_n\otimes\btau^{1/2} ]
\left[
\I_n \otimes\A
-
\left[
\matrix{
\0 & & \B\delta_{ij} \cr
& \ddots& \cr
\B^\prime\delta_{ij} & & \0
}
\right]
\right]^{-1}
[\I_n\otimes\btau^{1/2} ],
\end{equation}
where $\btau^{1/2} = [\tau_1,\ldots,\tau_p]^\prime$ and where $\{
\delta
_{ij}\}$ are indicator functions for neighborhood dependence.

The specifications above simply ensure symmetry and reduce the number
of parameters that must be estimated. Of course, the collection of
spatial-dependence parameters, $\{\rho_{j\ell}\}$ and $\{\phi_{j\ell
}\}
$, must be chosen to ensure that (\ref{cov2}) is a positive-definite
covariance matrix. The final model has $p(p-1)/2$ within-location
dependence parameters, $\{\rho_{j\ell}\}$, and $p^2$ between-location
spatial-dependence parameters, $\{\phi_{j\ell}\}$, in addition to the
$p$ variance parameters, $\{\tau_j^2\}$, and any parameters that are
used to define the means.


\section{A hierarchical model for an RCM experiment}
\label{sec:hier}

Let the $n$-dimensional vector $\mathbf{y}_{rj}$ denote
the output of an RCM,
in particular, the $r$th ensemble member for the $j$th variable. In
this work we focus solely on simple ensembles; that is, each member of
the ensemble represents a perturbation of initial conditions for a
single model. Potential extensions of this basic framework for
perturbed physics or multi-model ensembles is discussed in Section \ref
{sec:conclude}.

At the first level of the hierarchy, the data model assumes that the
vectors $\mathbf{y}_{rj}, r=1,\ldots,m, j=1,\ldots,p,$
are independent with
%
\begin{equation}\label{dm}
\mathbf{y}_{rj}|\balpha_j,\bbeta_{rj}, \h_{rj},\sigma
_j^2 \sim\cN(\X_1\balpha
_j+\X_2\bbeta_{rj} + \h_{rj}, \sigma_j\I), 
\end{equation}
where $m$ indicates the number of ensemble members. In the mean
structure, we allow for fixed effects common to all ensemble members
within the $j$th variable ($\X_1\balpha_j$) and random effects specific
to the $r$th ensemble member within the $j$th variable ($\X_2\bbeta
_{rj}$). Spatial random effects are included through $\h_{rj}$, and
$\sigma_j$ represents a variable-specific variance.

The process model has two parts. First, the vectors 
$[\bbeta_{r1}^\prime,\ldots,\bbeta_{rp}^\prime]^\prime,
r=1,\ldots,m,$
are assumed to be independent with
%
\begin{equation} \label{betaprior}
\pmatrix{
\bbeta_{r1} \cr
\vdots\cr
\bbeta_{rp}
}
\left|
\pmatrix{
\bbeta_1 \cr
\vdots\cr
\bbeta_p
}
,\right.
\qquad \bSigma_b
\sim
\cN\left(
\pmatrix{
\bbeta_1 \cr
\vdots\cr
\bbeta_p
},
\bSigma_b
\right),
\end{equation}
where $\bSigma_b$ is a $pq \times pq$ covariance matrix with $q$ the
number of columns of $\X_2$.
Second, the vectors 
$[\h_{r1}^\prime,\ldots,\h_{rp}^\prime]^\prime, r=1,\ldots,m,$
are\vadjust{\goodbreak}
assumed to be independent with
%
\begin{eqnarray}\label{hprior}
&&\pmatrix{
\h_{r1} \cr
\vdots\cr
\h_{rp}
}
\left|
\pmatrix{
\h_1 \cr
\vdots\cr
\h_p
}
,\right.\nonumber\\[-8pt]\\[-8pt]
&&\{\tau_j^2\},\{\rho_{j\ell}\}, \{\phi_{j\ell}\}
\sim
\cN\left(
\pmatrix{
\h_1 \cr
\vdots\cr
\h_p
},
\V(\{\tau_j^2\},\{\rho_{j\ell}\}, \{\phi_{j\ell}\} )
\right).\nonumber
\end{eqnarray}
(Note that the vectors $[\bbeta_{r1}^\prime,\ldots,\bbeta
_{rp}^\prime
]^\prime$ and $[\h_{r1}^\prime,\ldots,\h_{rp}^\prime]^\prime$ are
assumed to be independent as well.)
The first part, (\ref{betaprior}), focuses on linking the random
regression coefficients specific to each ensemble member, while the
second part, (\ref{hprior}), imposes a multivariate structure on the
spatial random effects. The covariance matrix $\V$ takes its form from
the multivariate Markov random field in (\ref{cov2}).

The final level of the hierarchy assumes prior distributions on $\{
\sigma_{j}\}$, $\{\balpha_j\}$, $\{\bbeta_j\}$, $\{\h_j\}$,
$\bSigma
_{b}$, and the parameters of the spatial covariance, namely, $\{\tau
_j^2\}$, $\{\rho_{j\ell}\}$, and $\{\phi_{j\ell}\}$. Typically, these
priors will be vague or noninformative as well as independent. In
addition, the prior distribution on $\{\rho_{j\ell}\}$ and $\{\phi
_{j\ell}\}$ must ensure that the resulting covariance matrix is
positive-definite.

From Bayes' theorem, the posterior distribution for the three-level
hierarchical model is given by
\begin{eqnarray*}
&&P(\{\bbeta_{rj}\},\{\h_{rj}\},\{\balpha_j\},\{\sigma_j\}
,\{
\bbeta_j\},\bSigma_{b},\{\h_j\},\{\tau_j^2\},\{\rho_{j\ell}\},\{
\phi
_{j\ell}\}|\Y)\\
&&\qquad \propto
P(\Y|\{\balpha_j\},\{\bbeta_{rj}\},\{\h_{rj}\},\{\sigma_j\}) \\
&&\quad \qquad {}\times
P(\{\bbeta_{rj}\}|\{\bbeta_j\},\bSigma_{b})
P(\{\h_{rj}\}|\{\h_j\},\{\tau_j^2\},\{\rho_{j\ell}\},\{\phi
_{j\ell}\})
\\
&&\quad \qquad {}\times
P(\{\balpha_j\})P(\{\sigma_{j}\})P(\{\bbeta_j\})P(\bSigma_{b})P(\{
\h_j\}
)P(\{\tau_j^2\})P(\{\rho_{j\ell}\},\{\phi_{j\ell}\}).
\end{eqnarray*}
It is clear that there is no closed-form solution for the posterior,
and here Markov chain Monte Carlo (MCMC) [e.g., \citet{Gilks96}] is
used to simulate realizations from the posterior distribution. In
particular, we implement a Gibbs sampler [\citet{GemGem84}; \citet{GelfSmith90}; \citet{Gelf90}], incorporating Metropolis--Hastings steps
[\citet{Metro53}; \citet{Hast70}] where necessary.

One benefit of a MRF is that the specification involves the precision
or inverse covariance matrix, and this matrix is typically sparse; that
is, many of the elements of the matrix are zero. Methods for storing
and manipulating such matrices have been widely established [e.g.,
\citet
{Davis06}], and there is great potential for computational efficiency
associated with sparse-matrix methods. There are now several
sparse-matrix packages in the R statistical computing environment
[\citet
{R}]. However, the \texttt{spam} package [\citet{spam}] has functionality
that is well suited for implementing MCMC with a MRF model. For
example, the sparse Cholesky decomposition is one of the most important
computational devices used when implementing the Gibbs sampler for MRF
models such as those developed in this work. A typical sparse Cholesky
decomposition involves three steps: (1) reorganizing the matrix by
permuting the rows/columns to achieve a pattern of sparsity that is
more efficient for the sparse Cholesky algorithm; (2) a symbolic step
that identifies the pattern of sparsity in the matrix; and (3) the
numerical computation. The first two steps do not change when
manipulating matrices repeatedly with the same patterns of sparsity (as
is the case here). The \texttt{spam} package allows one to achieve even
greater computational efficiency by not repeating these steps during
the course of the MCMC. For more details on this and other
computational benefits gained from incorporating sparse-matrix methods
in such applications, see \citet{FurrerSain2010}.

\section{The RCM experiment}
\label{sec:ex}

Leung et al. (\citeyear{Leung04}) describe an RCM experiment using the NCAR/DOE Parallel
Climate Model to drive the NCAR/Penn State Mesoscale Model (MM5) as an
RCM. The experiment produced a control run from 1995--2015 and three
future runs (ensemble members) from 2040--2060. The domain consisted of
the western United States and part of western Canada, and the model
used a ``business as usual'' climate scenario incorporating a 1\%
annual increase in the amount of greenhouse gases.

\begin{figure}

\includegraphics{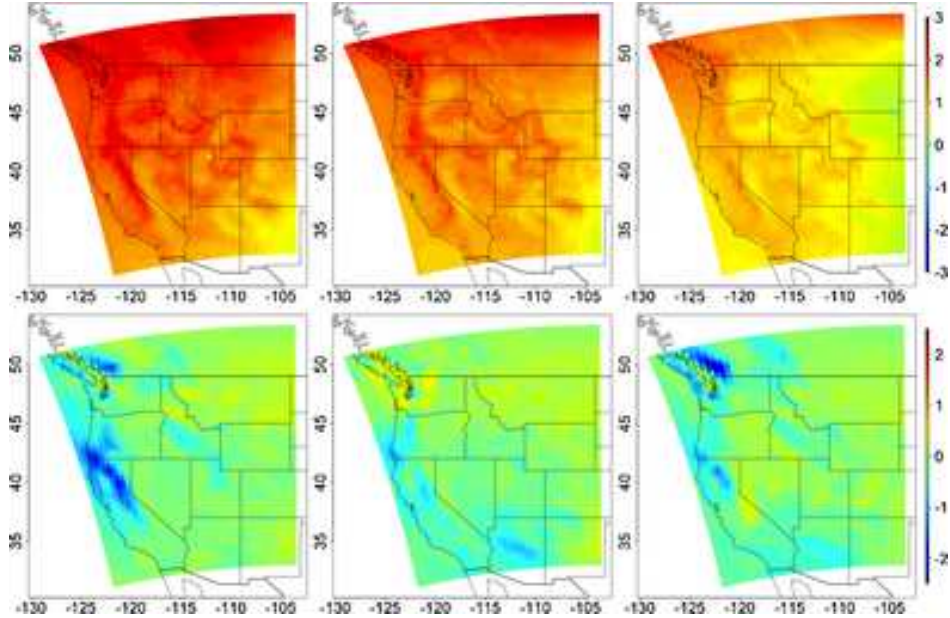}

\caption{Top row shows the three differences in winter midpoint
temperature ($^{\circ}$K), while the bottom row shows the three differences
in total precipitation (inches).}
\label{data}
\end{figure}

The $n=44 \times56 = 2464$ grid boxes form a regular lattice. Since a
long-run average of weather is one way to quantify climate, typical
summaries of climate model runs include seasonal averages of
temperature and precipitation with the length of the integration often
determined by computational considerations. Hence, twenty-year winter
(December, January, and February) average temperature and average total
precipitation were computed for each grid box and for each of the
control and the three future runs.
Differences between the future and the control were calculated,
yielding change-in-temperature and change-in-total-precipitation variables.
Hence, there are $p=2$ variables and $m=3$ ensemble members, giving six
fields to be analyzed. These spatial fields for the winter season are
shown in Figure \ref{data}. A second, separate analysis with the same
structure is also presented for the twenty-year summer (June, July, and
August) change in temperature and change in total precipitation.

We note that the statistical models outlined in this work focus on
using Gaussian assumptions for average precipitation from an RCM. (An
assumption that was verified through exploratory analysis.) However,
other models for precipitation, in particular, for extreme
precipitation, are possible; see, for example, \citet{Sanso2004},
\citet{Schliep2009}, and \citet{Cooley2010}.

\subsection{Model specification}

We now outline some specifics about the statistical-model
specification. Consider the data model (\ref{dm}). After some
exploratory analysis, (scaled) latitude, longitude, and elevation were
used as covariates in the common regression component ($\X_1\balpha
_j$), to which a random intercept across ensemble members was added
($\X
_2\bbeta_{rj}$). 
The prior covariance matrix for the random intercept in (\ref
{betaprior}) was also simplified to $\bSigma_b = \sigma_b^2\I_p$.

The prior distribution for the variance parameters, $\{\sigma_j^2\}$
and $\sigma^2_b$, were taken to be noninformative; that is, they were
assumed to follow the prior distribution $P(\sigma^2) \propto1/\sigma
^2$, independently. The prior distributions for the regression
parameters, $\{\balpha_j\}$ and $\{\bbeta_j\}$, were taken to be
mean-zero Gaussian distributions with covariance matrices proportional
to the identity and with large variances (i.e., $\sigma^2_{\balpha}=10$
and $\sigma^2_{\bbeta}=100$). The prior distributions for $\{\h_j\}$
were also taken to be mean-zero Gaussian distributions with covariance
matrices proportional to the identity and with large variances (i.e.,
\mbox{$\sigma^2_{\h_j} = 10$)}.

Finally, the prior specification for the joint distribution of $\rho$,
$\phi_{11}$, $\phi_{22}$, $\phi_{12}$, and $\phi_{21}$ was taken to be
uniform over the range of values that yield a positive-definite
covariance matrix. This region was identified using rejection sampling
based on a sparse Cholesky decomposition. A simple simulation study of
a univariate Markov random field's spatial dependence parameter (not
reported here) suggested that concentrating priors on a subregion of
the parameter space leads to a biased estimate when the true parameter
lies outside this region. While this may not seem surprising, the
lesson learned is that there has to be a good reason to choose
nonuniform priors for these bounded spatial-dependence parameters.

\subsection{Results for the winter season}

Posterior distributions were obtained using MCMC algorithms, and
considerable care was taken to ensure the convergence of the parameters
in the MCMC. This is especially true with respect to the
conditional-dependence parameters, where our experience has shown that
straightforward approaches can lead to disappointing performance (i.e.,
very slow mixing and convergence).
Ten chains were run, each with random starting values; the starting
values for the conditional-dependence parameters were chosen uniformly
across the space of values that yield a positive-definite covariance matrix.

A Gibbs sampler was implemented that involved three distinct regimes.
In the first regime (2500 iterations), each of the
conditional-dependence parameters was updated one at a time using a
Metropolis--Hastings algorithm. Gaussian proposal distributions were
used, with periodic updates of the proposal variance to achieve an
approximate 20\% acceptance rate. In the second regime (the next 10,000
iterations), $\rho$, $\phi_{12}$, and $\phi_{21}$ were updated
simultaneously using a Metropolis--Hastings algorithm with a
multivariate Gaussian proposal distribution. Again, the proposal
covariance matrix was updated periodically to achieve an approximate
20\% acceptance rate. Other conditional-dependence parameters were
still updated using a univariate Metropolis--Hastings algorithm.
Finally, in the third regime (the last 10,000 iterations), $\rho$,
$\phi
_{12}$, and $\phi_{21}$ were again updated simultaneously, but no
further updates of the proposal distribution were made. Convergence of
the posterior distributions of the parameters in the MCMC was monitored
using both graphical and numerical methods [e.g., \citet{Gelm96}].
Posterior distributions were then estimated by sampling from the third regime.

\begin{figure}[b]

\includegraphics{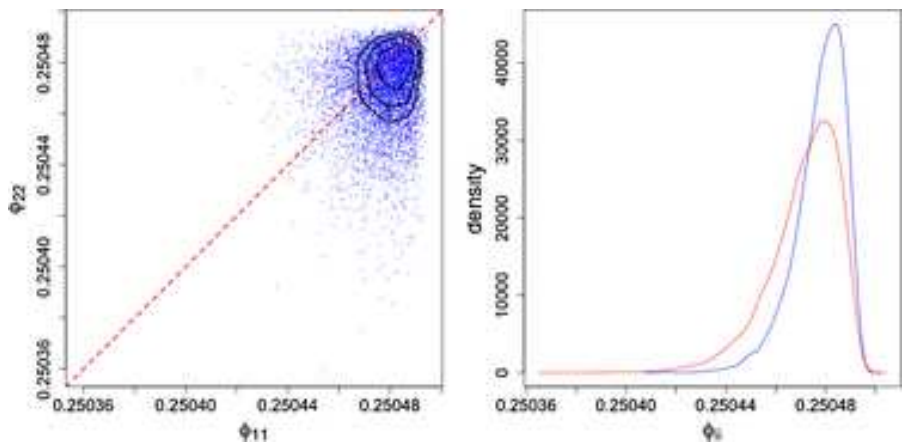}

\caption{Left frame shows scatterplot of a random sample of 10,000
values of $\phi_{11}$ and $\phi_{22}$ (1000 from each of the 10
chains). Contours represent approximate 25, 50, and 75\% contours of a
kernel density estimate. Right frame shows kernel density estimates of
the marginals for $\phi_{11}$ (blue) and $\phi_{22}$ (red). }
\label{cp1}
\end{figure}

Of particular interest are the conditional-dependence parameters, since
these control the nature and degree of the spatial correlation in the
model. Figure \ref{cp1} shows scatterplots and kernel estimates of the
distribution of $\phi_{11}$ (temperature) and $\phi_{22}$
(precipitation), the parameters that control the conditional dependence
between lattice points within a layer (Figure \ref{neigh}, left panel).
The distributions show that there is considerable (conditional) spatial
dependence within each variable, as the distributions tend to be
concentrated near the positive boundary of possible values for $\phi
_{11}$ and $\phi_{22}$. There is evidence of a slightly stronger
dependence for temperature ($\phi_{11}$).


\begin{figure}[t]

\includegraphics{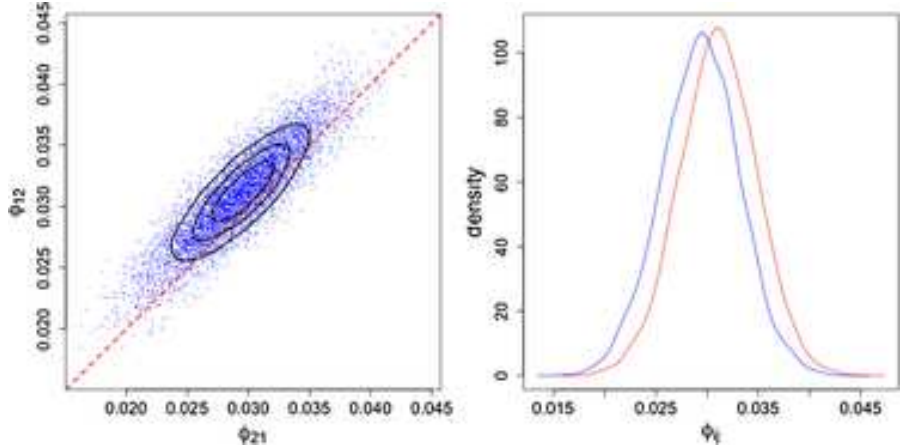}

\caption{Left frame shows scatterplot of a random sample of 10,000
values of $\phi_{12}$ and $\phi_{21}$ (1000 from each of the 10
chains). Contours represent approximate 25, 50, and 75\% contours of a
kernel density estimate. Right frame shows kernel density estimates of
the marginals for $\phi_{12}$ (red) and $\phi_{21}$ (blue). }
\label{cp3}
\end{figure}

Trace plots and other diagnostics for $\rho$, $\phi_{12}$, and $\phi
_{21}$ suggest convergence after about 10,000 iterations, which
corresponds to the end of the second sampling regime. These three
parameters control the dependence structure across variables; $\rho$
summarizes the within-location dependence (Figure \ref{neigh}, middle
panel) and $\phi_{12}$, $\phi_{21}$ summarize the cross-variable
dependence (Figure \ref{neigh}, right panel). The estimated posterior
mean and posterior standard deviation for $\rho$ is $-0.12$ and 0.014,
respectively. A negative value for $\rho$ suggests that an increasing
temperature is (conditionally) associated with a decreasing total precipitation.

Figure \ref{cp3} highlights the distribution of $\phi_{12}$ and $\phi
_{21}$. The strong correlation between these two conditional
cross-correlation parameters is clearly shown in the left frame of
Figure \ref{cp3}. However, there is another feature of note: There is
compelling evidence of asymmetry in the strength of these two
parameters, with roughly 85\% of the sampled points being above the
line $y=x$. This suggests that there is higher conditional dependence
between temperature values and neighboring total precipitation than
there is conditional dependence between total precipitation values and
neighboring temperature values.
Almost all of the published models of multivariate MRFs assume $\phi
_{12}=\phi_{21}$, something we have argued previously as being overly
restrictive [\citet{SainCres07}]. Our posterior inference shows the
inappropriateness of such an assumption in this case.

\begin{figure}

\includegraphics{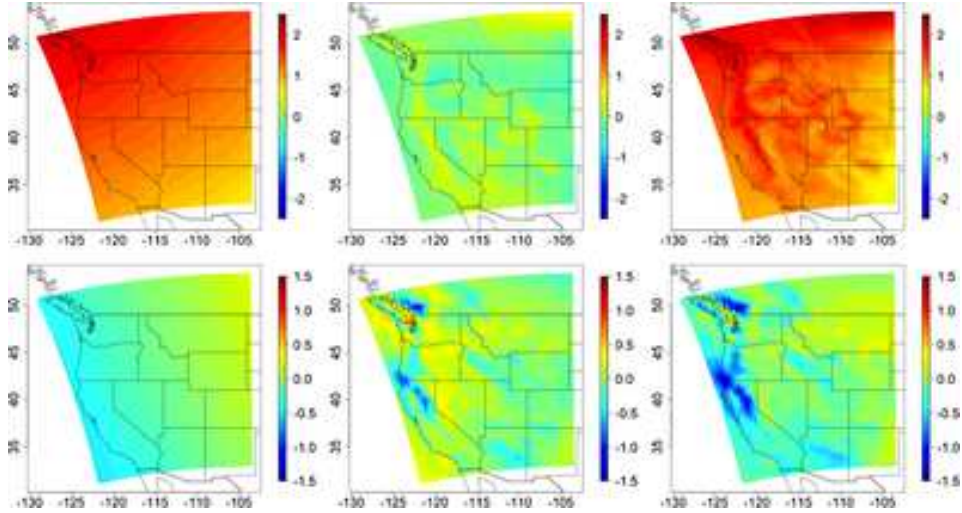}

\caption{Posterior means for the regression \textup{(left)}, the spatial effect
(middle), and the sum \textup{(right)} for the winter season. The top row
represents the change in midpoint temperature ($^{\circ}$K), while the
bottom represents the change in total precipitation (inches).}
\label{pm}
\end{figure}

Figure \ref{pm} shows posterior means for the fixed regression
components (left column), the spatial random effects (middle column),
and their sum (right column), for the change in winter average
temperature (top row) and the change in total winter precipitation
(bottom row). The fixed effects show a clear latitudinal effect as well
as an east-to-west gradient. For precipitation, there is a more
dominant east-to-west gradient. The spatial random effects for the
change in temperature seem to follow the features of the topography,
and are, in general, of smaller magnitude than the fixed effects. The
spatial effects for the change in total precipitation also follow the
features of the topography, but there are additional strong local
features, for example, in northern California. In contrast to
temperature, the spatial effects for total precipitation are larger
relative to the fixed effects.

The sum of the fixed effects and the spatial random effects for
temperature shows a consistent pattern of winter warming on average
throughout the west, while the sum for total precipitation shows
patterns that are much more localized. The most dominant signal for
total precipitation is indicated by the regions of sharp decline in
winter precipitation in northern California and the Pacific northwest.

\begin{figure}

\includegraphics{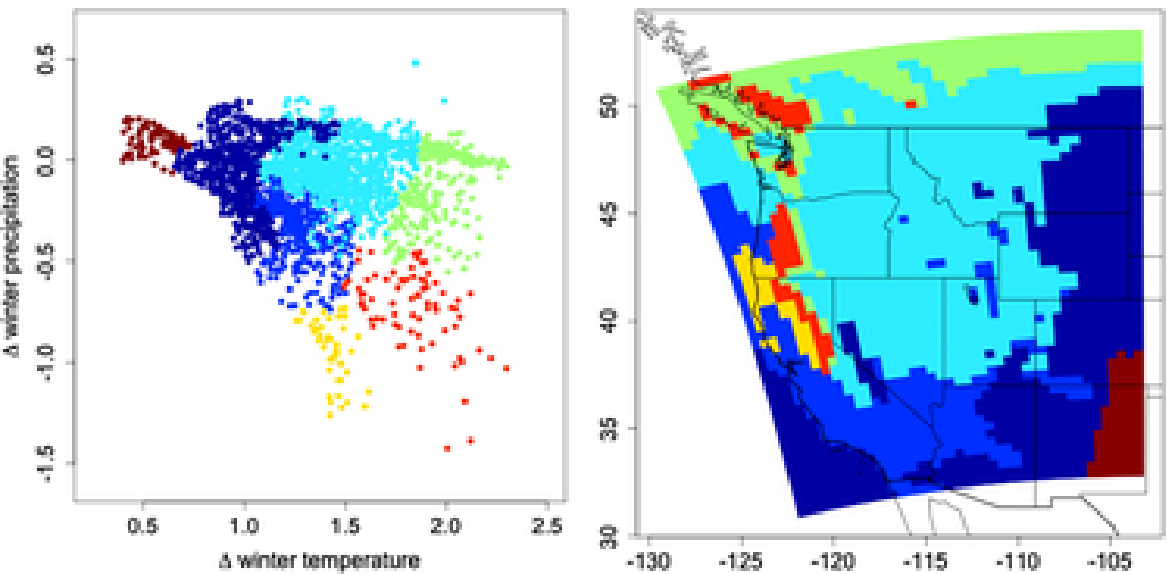}

\caption{Results from clustering the posterior means of the change in
temperature and the change in total precipitation for the winter
season. The left frame is a scatteplot with the clusters indicated
through different colors. The right frame shows the clusters spatially.}
\label{cl}
\end{figure}

To aid in the identification of areas that might be at most risk for
change, as projected by this regional-climate-model experiment, Figure
\ref{cl} shows the result of a hierarchical clustering based on the
posterior distribution of the mean change in temperature and total
precipitation for each grid. One thousand samples were drawn from the
posterior distribution of the mean change in temperature and total
precipitation for each grid box. The distance metric used to join
clusters was a symmetrized version of the Kullback--Leibler distance,
which was based on the assumption of bivariate normality within each
cluster. Hence, the clustering focuses not only on the mean of the
posterior distribution but also includes information about the
covariance structure of the changes for each grid box. The scatterplot
in the left frame of Figure \ref{cl} shows the posterior means for each
grid box with each cluster indicated by different colors. The
scatterplot shows the considerable structure that the clustering is
able to distinguish. The spatial pattern of the clusters is also shown
in Figure \ref{cl} (right frame). The dark red areas, for example,
highlight a region associated with a strong increase in temperature and
a strong decrease in total precipitation.

\begin{figure}[b]

\includegraphics{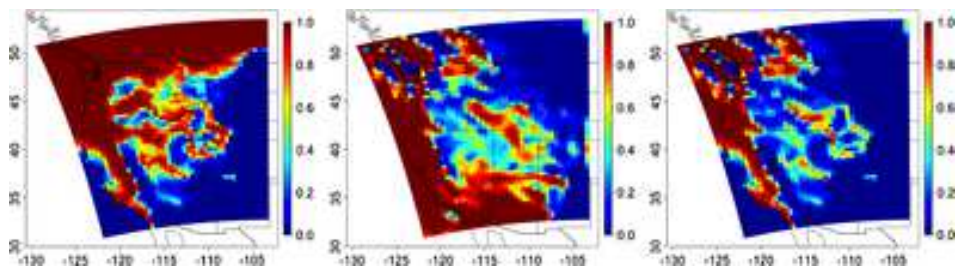}

\caption{Estimated pointwise probabilities for the winter season:
increasing temperature \textup{(left)}, decreasing total precipitation \textup{(middle)},
and simultaneously increasing temperature and decreasing total
precipitation \textup{(right)}.}
\label{probfig}
\end{figure}

It is also useful to consider other measures of uncertainty. Fields of
standard deviations are one approach, but when considering the
multivariate nature of this model output other possibilities arise. For
example, Figure \ref{probfig} shows estimated pointwise probabilities
of an increase in temperature (left frame), a decrease in total
precipitation (middle frame), and a simultaneous increase in
temperature and decrease in total precipitation (right frame) based on
a sampling of the posterior distribution of the joint spatial fields.
In general, temperature is increasing on average across the entire
domain; hence, these probabilities are based on increases larger than
the median computed from all the samples across all the grid boxes.
Likewise, the decrease in total precipitation was based on decreases
larger than the median across all the samples from all the grid boxes.
Again, on the basis of this model, we see evidence of widespread
increase in temperature and decrease in total precipitation across the
western U.S., with dominant features along the far western coast, the
Pacific northwest, and isolated mountain regions.

\begin{figure}[b]

\includegraphics{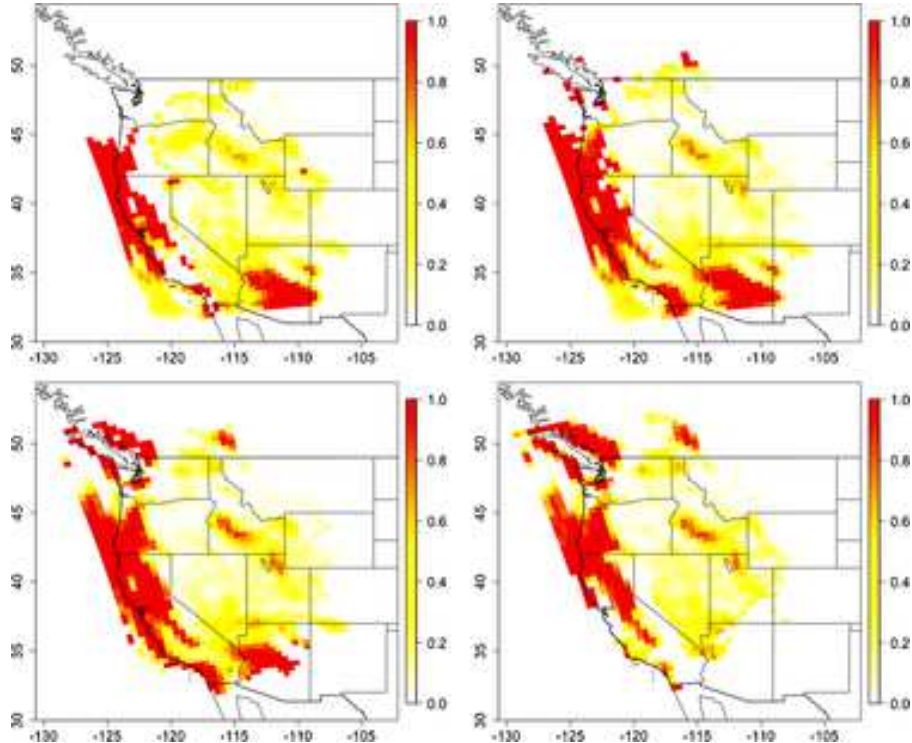}

\caption{Probability of a large decrease in winter total precipitation,
conditional on the increase in temperature falling in the first
quartile \textup{(top left)}, second quartile \textup{(top right)}, third quartile
\textup{(bottom left)}, and fourth quartile \textup{(bottom right)}.}
\label{winterconditional1}
%
%
%
%
%
%
%
%
\end{figure}

Figure \ref{winterconditional1} shows an alternative representation of
the joint distribution by considering conditional probabilities. 
The figure shows the probability of the decrease in total precipitation
being in the top quartile, conditional on the increase in temperature
being in the first quartile (top left), second quartile (top right),
third quartile (bottom left) and fourth quartile (bottom right). As the
temperature increase becomes more extreme, the largest decreases in
total precipitation move from being focused in the southwest (and the
California coast) to the Pacific northwest (and the California coast).
Relative changes can also be considered and, although not shown here,
the normalization minimizes the impact on the Pacific northwest (higher
absolute changes in total precipitation, but also higher total
precipitation values in general).


\begin{figure}

\includegraphics{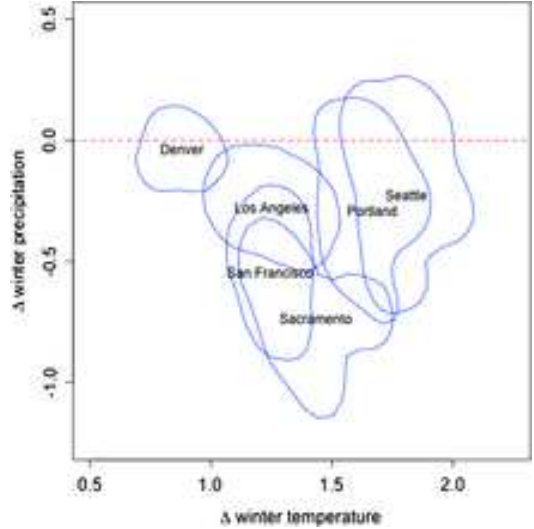}

\caption{Approximate 95\% contours for the average change in winter
temperature and precipitation for the grid boxes associated with the
five consolidated metropolitan statistical areas in the domain.}
\label{wintercontours}
\end{figure}

Finally, Figure \ref{wintercontours} shows approximate 95\% contours
for the joint change in (average winter) temperature and total
precipitation, but focuses on the grid boxes that represent the five
consolidated metropolitan statistical areas (as defined by the U.S.
Census) that are included in the domain. Figure \ref{wintercontours}
suggests that the projection for the Denver area includes average
increases in temperature of just under $1^\circ $K with minimal average
decreases in total precipitation. The contour for the Sacramento area,
on the other hand, suggests much larger average increases in
temperature and average decreases in total precipitation. We believe
that such plots, summarizing the joint distribution based on the
statistical model, will have great interest and application to
scientists and decision-makers interested in the impacts of climate change.

\subsection{Results for the summer season}

A slightly different scheme was used for the Gibbs sampler for the
analysis of the summer model output. The three-regime sampling was
still used, although with twice as many iterations (20,000) in the
second regime. In addition, all five conditional-dependence parameters
($\rho$, $\phi_{11}$, $\phi_{22}$, $\phi_{12}$, and $\phi_{21}$) were
updated simultaneously. Convergence of parameters for the summer season
was similar to that for the winter season, although somewhat slower,
and the specifics of those results are not discussed here.
Distributions of the posteriors for the parameters were similar, with
the exception of the cross-dependece parameters, and, in particular,
$\rho$ (posterior mean of $-0.41$ for summer versus a posterior mean of
$-0.12$ for winter), suggesting that the summer season has a much
stronger and more negative correlation between the change in
temperature and the change in total precipitation.

\begin{figure}

\includegraphics{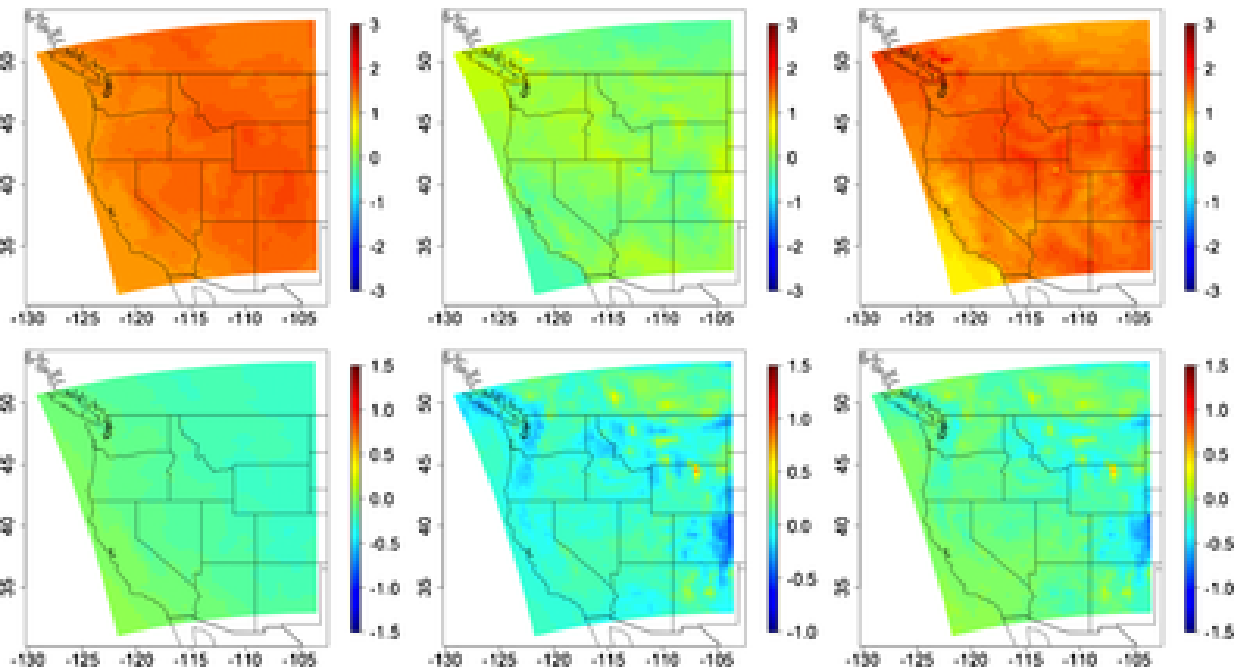}

\caption{Posterior means for the regression \textup{(left)}, the spatial effect
\textup{(middle)}, and the sum \textup{(right)} for the summer season. The top row
represents the change in midpoint temperature ($^{\circ}$K), while the
bottom represents the change in total precipitation (inches).}
\label{pm2}
\end{figure}

Figure \ref{pm2} shows posterior means for the summer season with the
same layout as in Figure \ref{pm} for the winter season. Now there
appears to be a west-to-east gradient in the fixed effects for
temperature, and, again, the spatial random effects pick up more of the
topography that is not accounted for in the fixed effects. Again, for
temperature, the spatial random effects are of smaller magnitude than
the fixed effects.

For total precipitation, there is also a west-to-east gradient in the
fixed effects. The spatial random effects for the change in total
precipitation also follow the features of the topography, but there are
strong local features, now occurring in the eastern part of the domain.
In comparison to temperature, the spatial random effects are larger
relative to the fixed effects.

\begin{figure}

\includegraphics{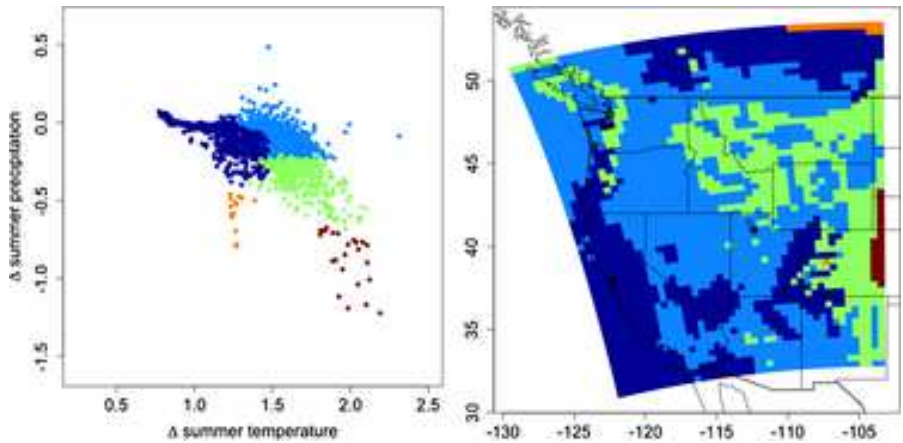}

\caption{Results from clustering the posterior means of the change in
temperature and total precipitation for the summer season. The left
frame is a scatterplot with the clusters indicated through different
colors. The right frame shows the clusters spatially.}
\label{cl2}
\end{figure}

The sum of the fixed effects and the spatial random effects for
temperature shows a consistent pattern of summer warming on average
throughout the west, while the sum for total precipitation shows
patterns that are much more localized, just as in the winter season.
However, the most dominant features for total precipitation are the
regions of decrease in summer total precipitation in the eastern part
of the domain. Again, a hierarchical clustering was performed on
samples from the posterior distribution for each grid box, which is
summarized in Figure \ref{cl2}. There appears to be more widespread
warming and decreasing total precipitation during the summer months,
but the clustering again highlights structure in the joint distribution.

%
%
%
%

\begin{figure}[b]

\includegraphics{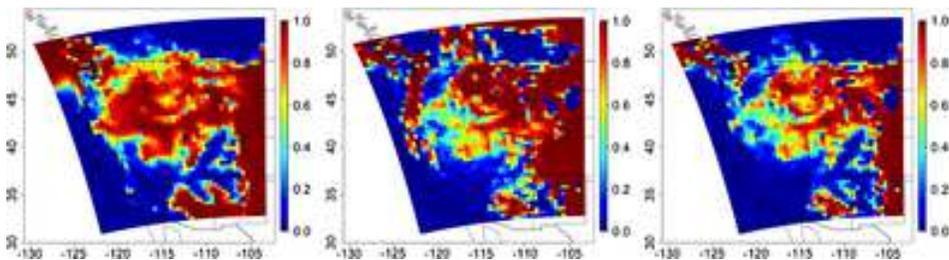}

\caption{Estimated pointwise probabilities for the summer season:
increasing temperature \textup{(left)}, decreasing total precipitation \textup{(middle)},
and simultaneously increasing temperature and decreasing total
precipitation \textup{(right)}.}
\label{probfig2}
\end{figure}

As in Figure \ref{probfig}, Figure \ref{probfig2} shows estimated
pointwise probabilities of an increase in temperature (left frame), a
decrease in total precipitation (middle frame), and a simultaneous
increase in temperature and decrease in total precipitation (right
frame) based on a sampling of the posterior distribution of the joint
spatial fields. In general, summer warming and decreasing total
precipitation is widespread, even more so than in the winter season,
and focused more on the eastern side of the domain.

%
%

\begin{figure}

\includegraphics{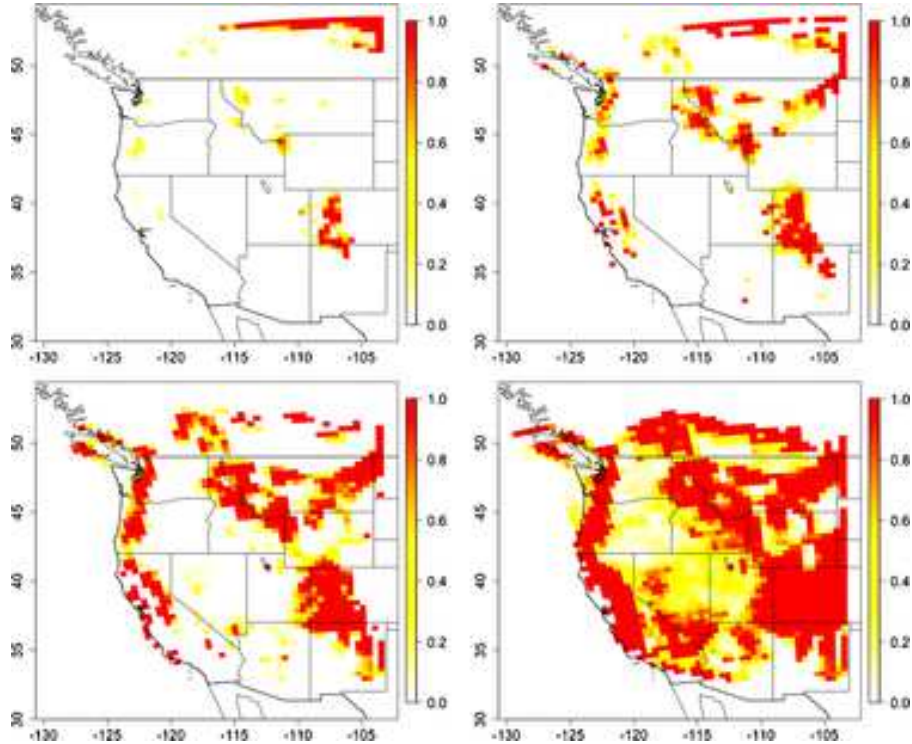}

\caption{Probability of a large decrease in summer total precipitation,
conditional on the increase in temperature falling in the first
quartile \textup{(top left)}, second quartile \textup{(top right)}, third quartile
\textup{(bottom left)}, and fourth quartile \textup{(bottom right)}.}
\label{summerconditional1}
%
%
%
%
%
%
%
%
\end{figure}

Figure \ref{summerconditional1} is constructed similarly to Figure
\ref{winterconditional1}.
However, the stronger negative correlation in the summer season is more
apparent in the figure. The large decreases in total precipitation are
strongest in the eastern portion of the domain, but decrease
dramatically when we condition on larger increases in temperature.

\begin{figure}

\includegraphics{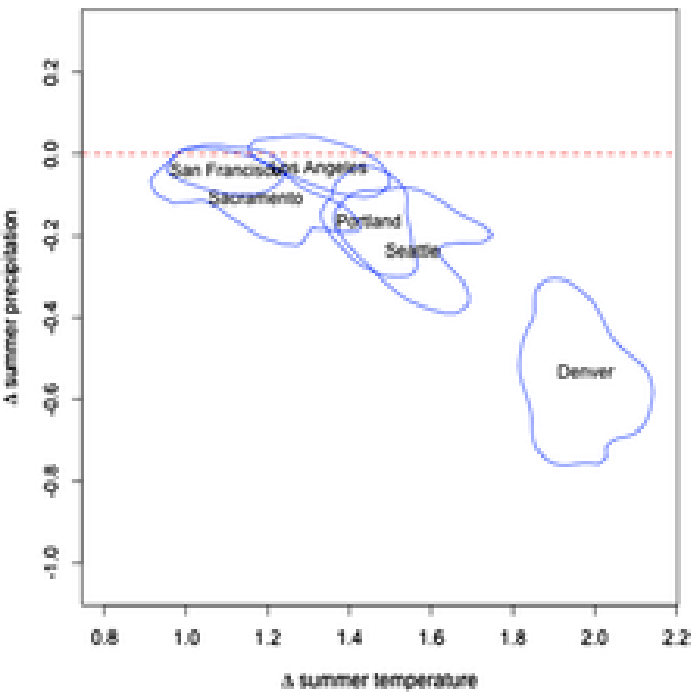}

\caption{Approximate 95\% contours for the average change in summer
temperature and precipitation for the grid boxes associated with the
five consolidated metropolitan statistical areas in the domain.}
\label{summercontours}
\end{figure}

Finally, Figure \ref{summercontours} shows approximate 95\% contours
for the joint change in (average summer) temperature and total
precipitation. In this case, the roles are reversed from the winter
season. The contours suggest larger increases in temperature and
decreases in total precipitation (on average) for the Denver area,
while the contour for the Sacramento area suggests more modest changes
on both variables.

\section{Concluding remarks}
\label{sec:conclude}

Climate models have become an important tool in the study of climate
and climate change. Ensemble experiments of climate-model output, be
they comprised of perturbed initial conditions, perturbed physics, or
multiple models, have also become important in studying and quantifying
the uncertainty in climate-model output. However, there are typically
only a limited number of runs that can be produced due to the time and
expense of running these models, even on modern supercomputers. Hence,
statistical methods become necessary to quantify the distribution and
the breadth of variation in the model output.

With this idea in mind, we have introduced a hierarchical spatial
statistical model designed primarily for the analysis of
regional-climate-model output on the basis of a simple ensemble
(perturbed initial conditions). This model is multivariate and has the
capacity to simultaneously characterize multiple model outputs, for
example, the average change in temperature and the average change in
total precipitation. While analysis of the individual model outputs
might yield estimates of marginal distributions, the strong
correlations across variables, such as those uncovered here in the
analysis of the average changes in summer temperature and the average
changes in total precipitation, make a multivariate analysis crucial
for joint inference.

The statistical model also captures the spatial variation in the model
output through a novel implementation of a multivariate MRF. In
addition to the computational benefits arising from using models based
on an MRF, this formulation of a multivariate MRF has a great deal of
flexibility in modeling the conditional-dependence structure and is
easily extendable. For example, more complex neighborhood structures
can be considered [e.g., \citet{sainfurrercressie2007}], and it is
not difficult to conceptualize how one might even consider modeling the
joint distribution of multiple variables that are on different
lattices. Connections to graphical models [e.g., \citet{Whittaker1990}]
could lead to further insights into modeling and parameter estimation.
The computational impact and practical utility of considering
additional variables (increasing $p$) or additional ensemble members
(increasing $m$) is also of interest, although, at least in climate
model research, these are highly dependent on the application and the
computational demands associated with running climate models on
high-performance computers. These are issues that we are currently
considering. We also note that this work adds to a growing collection
of research involving the study and modeling of asymmetric
cross-dependence structures for multivariate spatial data, including,
for example, \citet{Jin2005} and \citet{SainCres07} in the case of
MRFs and \citet{RoyleBerliner1999}, \citet{VerHoef2004}, and \citet{Apanasovich2010}
for geostatistical data.

There is great interest in more complex ensembles, such as
perturbed-physics experiments and multi-model ensembles. We have not
considered such ensembles here, as we have focused on the multivariate
aspect of the analysis of simple ensembles of regional-climate-model
output. However, the model presented here could be extended to consider
such ensembles by straightforward modifications to the process model,
in particular, equation (\ref{hprior}) could be modified to allow for a
spatial meta-analysis component [e.g., \citet{Kang2010}] or through a
functional analysis of variance similar to that of \citet
{kaufmansain2009}. Aside from the obvious computational challenges to
simply fitting such models in the multivariate setting, there is, of
course, more work needed to quantify the variation associated with
different model physics or different models.

It is important to note that any conclusions taken from an analysis
such as the one considered here are conditional on the assumptions
implicit to the particular climate model or models used to generate the
output fields. Whether global or regional in nature, climate models are
typically constructed to reproduce certain features in the current
climate, and analyzing differences as we did should minimize the impact
of any biases in the climate models (although this approach might be
viewed with some healthy skepticism, as it is not clear that the biases
in current runs are going to be the same as biases in future runs).
Observations may be included to help constrain the statistical model,
at least with respect to current climate. However, it is still an open
question how to include observational data sets for spatial analyses of
RCMs of the sort done here. Station-level data does not have the same
spatial and temporal coverage, and there are also numerous additional
issues with using interpolated data products, including reanalysis data
that represent a data assimilation using both station-level data and
climate-model output.

In addition to the longitude, latitude, and elevation used in this
analysis, predictors based on climatology (long-run means of
temperature and precipitation) were considered [e.g., \citeauthor{Fur07} (\citeyear{Fur07,furrersain2007})], but these were ultimately ruled out as not
being effective at predicting the changes in temperature and
precipitation. However, work done by \citet{Teb05} offers an approach
for combining model output and observations. We are currently
considering how their approach may be useful for spatial analyses of
RCM output.

Finally, there is also much interest, for example, from people
examining the impacts of climate change, in combining model output in
order to obtain improved projections of climate change or to span the
variation across a climate-model experiment. Of course, with the more
complex climate-model experiments, there is the issue of model-to-model
correlations. We believe that the inherent multivariate nature of this
model provides an excellent starting place to consider such correlations.

\section*{Acknowledgments}

We would like to express our appreciation to the referees and the
editor for their helpful comments.
The National Center for Atmospheric Research is sponsored by the
National Science Foundation (NSF).


\printaddresses

\end{document}